\documentclass[a4paper,aps,pre,showpacs,superscriptaddress,floatfix,nofootinbib,twocolumn,bibtex]{revtex4}

\usepackage{bbold}
\usepackage{bbm}
\usepackage[pdftex]{graphicx}
\usepackage{latexsym,amsmath,verbatim}
\usepackage{color}
\usepackage{rotating}
\usepackage{verbatim}
\usepackage{multirow}
\usepackage[english]{babel}
\usepackage{comment}

\begin{document}

\title{Predicting percolation thresholds in networks}

\author{Filippo Radicchi}
\affiliation{Center for Complex Networks and Systems Research, School of Informatics and Computing, Indiana University, Bloomington, USA}
\email{filiradi@indiana.edu.}

\date{\today}

\begin{abstract}
  \noindent
  We consider different methods, that do not
  rely on numerical simulations of the
  percolation process, to approximate
  percolation thresholds in networks.
  We perform a systematic analysis on
  synthetic graphs and
  a collection of $109$ real networks to
  quantify their effectiveness
  and reliability as prediction tools. 
  Our study reveals that the inverse of the largest eigenvalue
  of the non-backtracking matrix of the graph often 
  provides a tight lower bound for true
  percolation threshold. However, in more than $40\%$ of the cases,
  this indicator is less predictive than
  the naive expectation value based solely on the
  moments of the degree distribution.
  We find that the performance of all indicators becomes worse as
  the value of the true percolation threshold grows. Thus,
  none of them represents a good
  proxy for robustness of extremely fragile networks.
\end{abstract}

\pacs{89.75.Hc, 64.60.aq}

\maketitle

Percolation is one of the most studied
processes in statistical 
physics~\cite{stauffer1991introduction}.
The model assumes the presence of
an underlying network structure, where
nodes (site percolation) 
or edges (bond percolation) are 
independently occupied with probability 
$p$~\cite{albert2002statistical, dorogovtsev2008critical}. 
Nearest-neighboring occupied sites or bonds
form clusters.
For $p=0$, only clusters
of size one are present in the system.
For $p=1$ instead, a unique giant cluster
spans the entire network.
At intermediate values of $p$, the network can 
be found in two different phases:
the non-percolating regime, where
clusters have microscopic size, in the
sense that the number of nodes
within each cluster is much smaller than the
size of the network; the percolating phase, where
a single macroscopic cluster, whose size is comparable with the
one of the entire network, is present.
The value of $p$ that separates
the two phases is a 
network-dependent quantity 
called percolation threshold,
and it is usually denoted as $p_c$.
Percolation models are commonly used 
to study network robustness against random
failures~\cite{albert2000error, cohen2000resilience},
and the spreading of diseases
or ideas~\cite{pastor2001epidemic, kempe2003maximizing}.
In practical applications, knowing the value of 
the percolation threshold of a given network
is thus extremely important. 
For example, threshold values
of technological or infrastructural 
networks can be used to estimate the maximal number
of local failures that these systems 
can tolerate before stopping to function~\cite{holme2002attack}.
In the case of networks of social contacts, the
value of the percolation threshold can be
interpreted as a proxy for the risk to observe
disease outbreaks~\cite{meyers2007contact}. 
Although the importance of this feature, 
there are only a few methods, that do not rely
on direct numerical simulations
of the percolation process, able to
determine the value of the
percolation threshold in a
network~\cite{PhysRevLett.113.208702}.
In this paper, we consider three different indicators
that have been recently introduced
to provide estimates of bond percolation
thresholds in arbitrary networks.
We first measure their performances in synthetic graphs,
and then test their predictive power
in a large and heterogeneous set
of real-world networks.

\begin{figure}[!htb]
\includegraphics[width=0.48\textwidth]{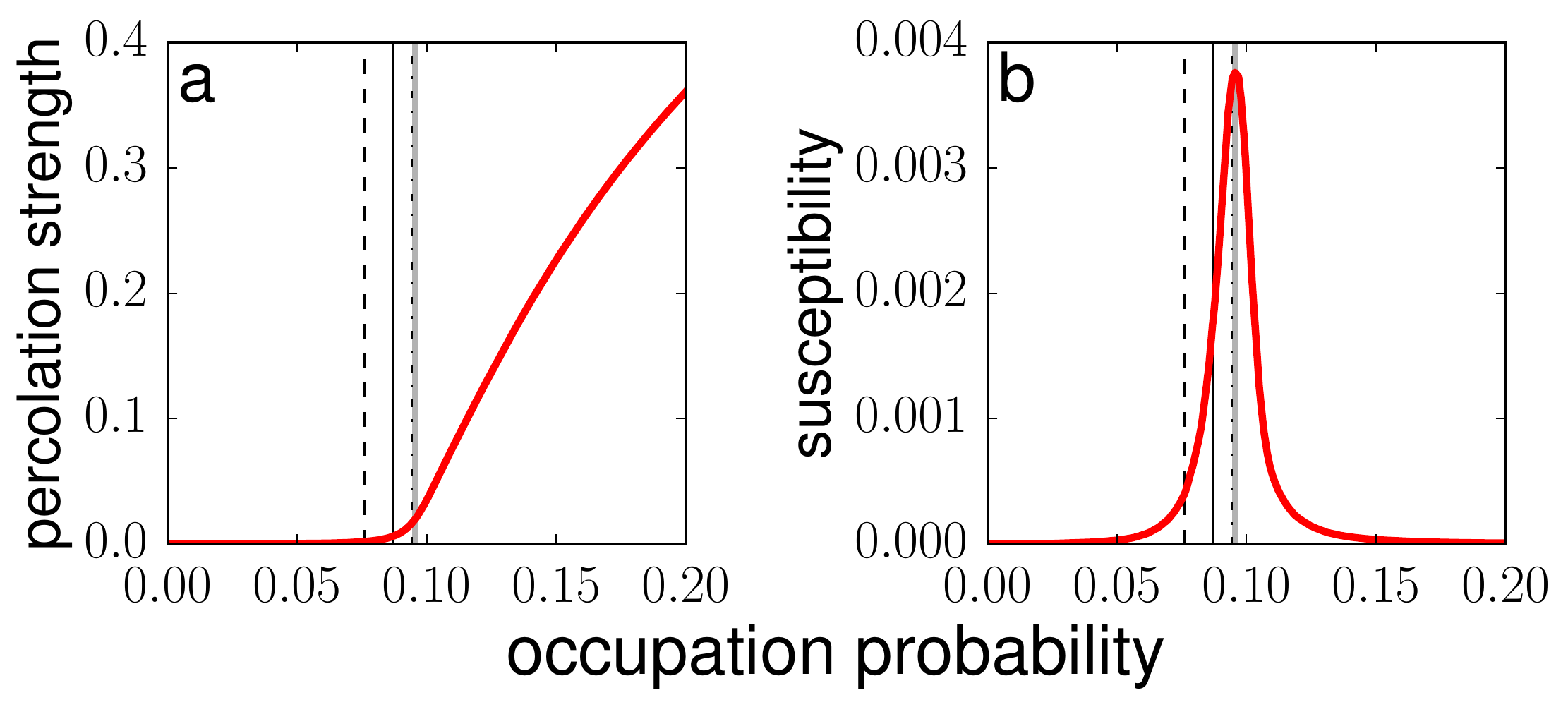}
\caption{(Color online) Bond percolation transition of the giant
connected component of the {\it peer-to-peer Gnutella}
network as of August 31, 
2002~\cite{leskovec2007graph, ripeanu2002mapping}. We consider
the undirected version of the network, originally directed.
The largest connected component
is composed of $N=62,561$ nodes and $E=147,878$.
a) Percolation strength $P_\infty$
as a function of the bond occupation 
probability $p$ [Eq.~(\ref{eq:perc})].
b) Susceptibility $\chi$ as a function of the
bond occupation probability $p$ [Eq.~(\ref{eq:susc})].
$P_\infty$ and $\chi$ are represented
as red thick lines and have been computed
with $Q=10,000$ independent realizations of the 
Montecarlo algorithm by Newman and Ziff~\cite{newman2000efficient}.
The best estimate of the percolation threshold 
$p_c = 0.0955$ [Eq.~(\ref{eq:critical}), full gray line] is compared
with the estimations $\tilde{p}_c = 0.0943$ [Eq.~(\ref{eq:moments}),
  dotted black line],
$\bar{p}_c = 0.0759$ [Eq.~(\ref{eq:adj}), dashed black line],
and $\hat{p}_c = 0.0871$
[Eq.~(\ref{eq:non}), black full line].
}
\label{fig:1}
\end{figure}
%

To this end, we first need to provide
the right term of comparison, i.e., 
the best estimate of the true threshold
by means of direct numerical simulations
of the percolation process. Given an undirected and unweighted 
network with $N$ nodes and $E$
edges composed of a single connected component,
we study bond percolation
using the Montecarlo method proposed by
Newman and Ziff~\cite{newman2000efficient}.
In each realization of the
method, we start from a configuration with
no connections. We then sequentially add
edges in random order, and monitor the evolution
of the size of the largest cluster
in the network $S(p)$ as a function
of the bond occupation probability $p = e/E$,
where $e$ indicates
the number of edges added from the initial
configuration, i.e., $e=0$. We repeat
the entire process $Q$ independent times,
and estimate the percolation strength as
\begin{equation}
  P_\infty (p) = \frac{1}{N\, Q} \; \sum_{q=1}^Q S_q(p)
\label{eq:perc}  
\end{equation}
and the susceptibility as
\begin{equation}
  \chi (p) = \frac{ \frac{1}{N^2\, Q} \; \sum_{q=1}^Q S_q(p)S_q(p) - \left[P_\infty (p)\right]^2 } {P_\infty (p)}\; ,
\label{eq:susc}  
\end{equation}
where $S_q(p)$ indicates the size of the
largest cluster in the network observed,
during the $q$-th realization of the Montecarlo
algorithm, when the bond occupation probability
equals $p$.
We determine the best
estimate of the percolation
threshold $p_c$ as the value of $p$
where the susceptibility reaches its maximum
\begin{equation}
  p_c = \arg{ \left\{ \max_p \, \chi (p) \right\} } \; .
\label{eq:critical}  
\end{equation}
A concrete example of the application
of the method described above to a real network
is provided in Fig.~\ref{fig:1}.
Note that Eq.~(\ref{eq:critical}) 
is not the only possible
way to compute the best estimate of the true percolation
threshold. Another very popular method is to
determine $p_c$ as the value of $p$
where the size of the second largest
cluster reaches its maximum~\cite{PhysRevLett.113.208702}.
Generally, the two definitions do not provide
the same exact value for the best estimates,
but values whose difference is within the
tolerance required
for the results of this paper.

As stated above, the main purpose of our paper 
is to understand the best way to predict the percolation
threshold of Eq.~(\ref{eq:critical}) without
performing direct numerical simulations.
We consider three different indicators.
Our first estimation is based on the quantity
\begin{equation}
  \tilde{p}_c =  \frac{\langle k \rangle}{ \langle k^2 \rangle - \langle k \rangle }\; .
\label{eq:moments}  
\end{equation}
In Eq.~(\ref{eq:moments}), $\langle k \rangle$
and $\langle k^2 \rangle$
respectively represent the first and
the second moments of the
degree distribution of the graph. $\tilde{p}_c$ represents
the value of the percolation threshold
expected in uncorrelated
graphs with prescribed degree
sequence~\cite{cohen2000resilience, callaway2000network}.
In general, such prediction should
fail in real networks where correlations are
present.
As a second prediction method of
the percolation threshold, we
consider the inverse of the largest
eigenvalue of the
adjacency matrix $A$ of the graph
\begin{equation}
  \bar{p}_c =  \left [ \max_{\vec{v}} \, \frac{\vec{v}^T \, A \, \vec{v}} {\vec{v}^T \, \vec{v}} \right]^{-1}\; .
\label{eq:adj}  
\end{equation}
This is the expected value of the percolation threshold
in graphs with a high density of connections~\cite{bollobas2010percolation}.
Given that real networks are generally sparse, we
expect therefore that this method is not able to
produce very good prediction values
of the percolation threshold.
Finally, we provide an estimation of the percolation
threshold as
\begin{equation}
  \hat{p}_c =  \left [ \max_{\vec{w}} \, \frac{\vec{w}^T \, M \, \vec{w}} {\vec{w}^T \, \vec{w}} \right]^{-1}\; ,
\label{eq:non}  
\end{equation}
where the matrix $M$ is defined as
\begin{equation}
  M = \left( \begin{array}{cc} A & \mathbbm{1} -D \\  \mathbbm{1} & \emptyset \end{array} \right) \;.
  \label{eq:non_matrix}
\end{equation}
$M$ is a square matrix
of dimension $2N \times 2N$ composed of four  $N \times N$ blocks.
$A$ is the adjacency matrix of the graph, $\mathbbm{1}$ is the identity
matrix, and $D$ is a diagonal matrix whose elements are
equal to the degree of the nodes. The largest eigenvalue
of $M$ is identical to the largest eigenvalue of the so-called
non-backtracking or Hasmimoto matrix associated with the 
graph~\cite{hashimoto1989zeta}. Spectral properties of this matrix
are relevant not
just in percolation theory, but also for clustering
algorithms
and centrality measures~\cite{krzakala2013spectral, martin2014localization}.
The indicator of Eq.~(\ref{eq:non}) is
obtained in the
approximation of locally tree-like networks, and it is
therefore expected to provide good predictions
for sparse networks~\cite{PhysRevLett.113.208701, PhysRevLett.113.208702}.

\begin{figure}[!htb]
\includegraphics[width=0.48\textwidth]{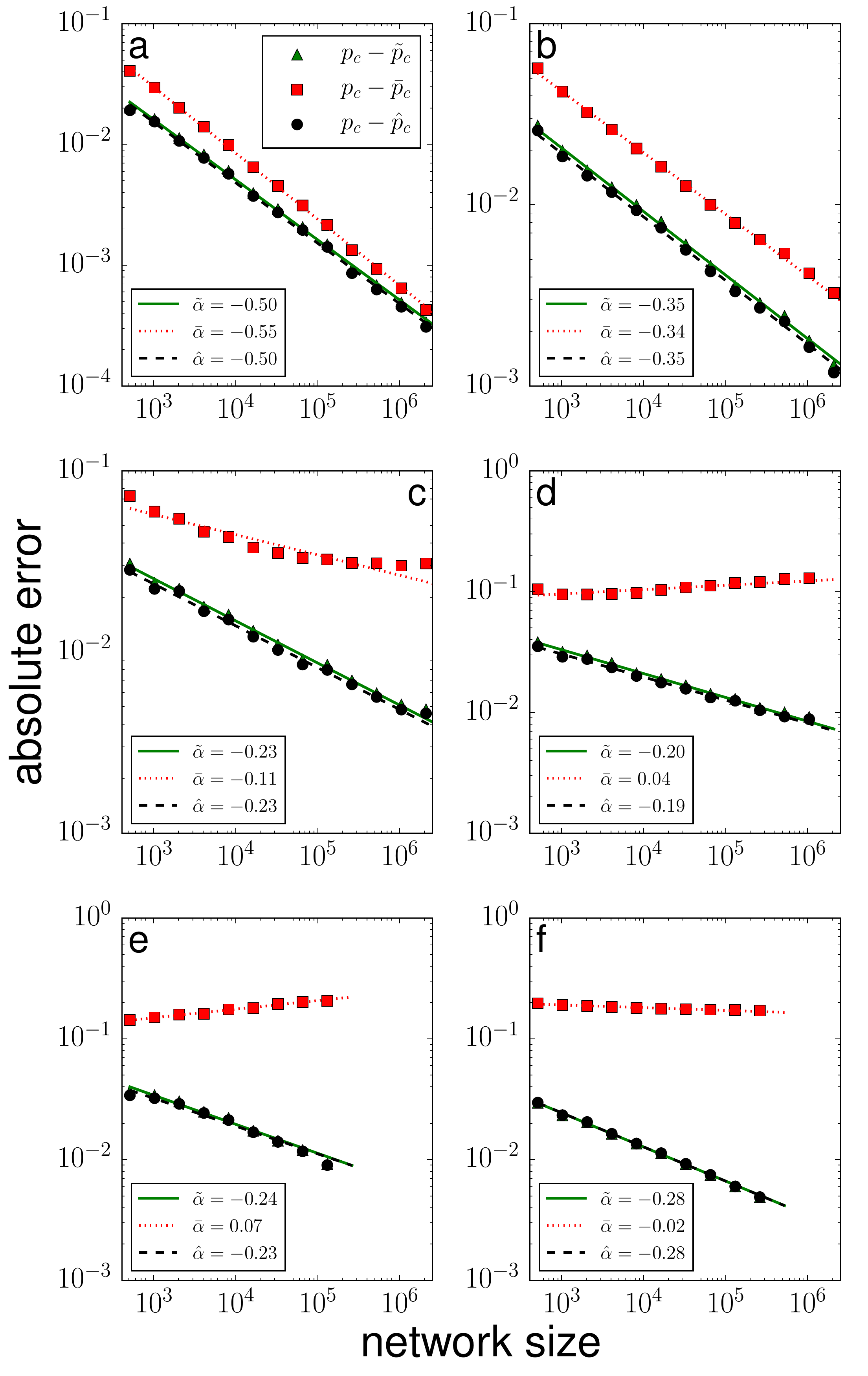}
\caption{
  (Color online) Difference between best estimates of the bond
  percolation threshold $p_c$ and the predicted
  values $\tilde{p}_c$ [Eq.~(\ref{eq:moments}), green triangles],
  $\bar{p}_c$ [Eq.~(\ref{eq:adj}), red squares],
  and $\hat{p}_c$ [Eq.~(\ref{eq:non}), black circles]
  in the uncorrelated configuration model~\cite{catanzaro2005generation}.
  Gaps between best estimates and predictions are
  plotted as function of the network size $N$.
  We report results obtained for networks
  generated according to the power-law
  degree distribution $P(k) \sim k^{-\gamma}$,
  with support $[3, \sqrt{N}]$. Different panels
  correspond to different choices of the degree exponent $\gamma$: 
  a) $\gamma = 2.1$, b) $\gamma = 2.5$, c) $\gamma = 2.9$, d)
  $\gamma = 3.5$, e) $\gamma = 4.5$, f)
  $\gamma = 100.0$. 
  Points are obtained by averaging over at least $50$ independent
  network instances. Standard errors on the measures have
  size smaller than those of the symbols.
  For each network, the best estimate of
  the bond percolation threshold is obtained with $Q=100$
  Montecarlo realizations of the Newman-Ziff
  algorithm~\cite{newman2000efficient}.
  In the various panels, 
  lines correspond to best fits of the empirical points
  with the functions $p_c - \tilde{p}_c \sim N^{\tilde{\alpha}}$ (full green
  lines),
  $p_c - \bar{p}_c \sim N^{\bar{\alpha}}$ (red dotted lines), and
  $p_c - \hat{p}_c \sim N^{\hat{\alpha}}$ (black dashed lines).
}
\label{fig:2}
\end{figure}

First, we perform a systematic study
on synthetic networks.
Each network realization is obtained with the use of the
so-called uncorrelated configuration model~\cite{catanzaro2005generation}.
This is a variation of the model by Molloy and Reed
to generate simple (i.e., without multiple or self
connections) random graphs with arbitrary degree
distributions~\cite{molloy1995critical}.
In our numerical experiments, we extract 
degrees from the power-law probability distribution
$P(k) \sim k^{-\gamma}$ if $k \in [3, \sqrt{N}]$, and 
$P(k) =0$, otherwise. Setting the degree cutoff
at $\sqrt{N}$ ensures the absence of degree-degree
correlations~\cite{boguna2004cut}, and the correct behavior of the
moments of the degree distribution~\cite{PhysRevE.90.050801}.
We consider six different values of
the degree exponent $\gamma$, ranging from $2.1$ to $100.0$,
and study the performance of
$\tilde{p}_c$, $\bar{p}_c$ and $\hat{p}_c$ as functions
of the network size $N$ (see Fig.~\ref{fig:2}).
We note that for every network instance, all 
these quantities always provide a value
smaller than $p_c$. We quantify the performance 
of a given predictor by measuring the difference 
between predicted value and best estimate $p_c$.
The inverse of the largest
eigenvalue of the adjacency matrix $\bar{p}_c$
provides reasonable predictions for the
percolation threshold only for values
of $\gamma < 3$. In that regime, the gap
$p_c - \bar{p}_c$ goes to zero, as $N$ increases,
in a power-law fashion. For larger values of the degree
exponent instead, the gap stabilizes to
finite values even in the limit
of infinitely large networks.
$\tilde{p}_c$ and $\hat{p}_c$ have
instead very good performances
for any value of $\gamma$. 
Their gaps with respect to $p_c$ always tend to zero
as the system size grows.
The scaling of the gaps is well fitted by
power-law functions having similar
exponent values. Although $\hat{p}_c$ is always closer 
to $p_c$ than $\tilde{p}_c$, the two measures essentially provide
equivalent estimates of the percolation threshold
on sparse random graphs.
This type of results can be understood
with an intuitive mathematical argument, according
to which we rewrite the
eigenvalue problem for the matrix $M$ of
Eq.~(\ref{eq:non_matrix}) as
$\left(\vec{w}_1^T, \vec{w}_2^T\right) M^T = \lambda \left(\vec{w}_1^T, \vec{w}_2^T\right)$, where $\vec{w}_1$ and $\vec{w}_2$
respectively contain the first and the second $N$
components of the eigenvector $\vec{w}$.
We can now write the eigenvalue
problem as two coupled equations:
$A \vec{w}_1 + (\mathbbm{1} - D) \vec{w}_2 = \lambda \vec{w}_1$
and $\vec{w}_1 = \lambda \vec{w}_2$.
Multiplying the first equation by the
row vector $\vec{u}^T = (1, \ldots, 1)$,
we obtain
\begin{equation}
  \lambda = \frac{\vec{d}^T \vec{w}_1}{\vec{u}^T \vec{w}_1} - 1 \; ,
  \label{eq:non_eig}
\end{equation}
where the components of the vector $\vec{d}$
are equal to the node degrees. In a random uncorrelated network,
we should expect the non-backtracking centrality
of the nodes be proportional to their degrees, 
i.e., $\vec{w}_1 \sim \vec{d}$,
and so Eq.~(\ref{eq:non}) reduces to
Eq.~(\ref{eq:moments})~\cite{martin2014localization}.

Second and more importantly, we perform a systematic
study of the predictive power of $\tilde{p}_c$,
$\bar{p}_c$ and $\hat{p}_c$ in a collection of
$109$ real networks. We consider graphs
of heterogeneous nature, including biological, infrastructural,
information, technological, social, and communication
networks, and thus very variegate also in
terms of structural properties (e.g., degree distribution
and correlations,
clustering coefficient, diameter).
In our numerical study, we reduce, if necessary, weighted and/or
directed networks to their unweighted and undirected
projections. Also, we focus our
attention only to their giant connected components.
Results for all $109$ real networks are provided in the
Supplemental Material. In Fig.~\ref{fig:3}, we summarize
the main outcome of our analysis.
As expected,
both $\bar{p}_c$ and $\hat{p}_c$ always
provide a lower bound for $p_c$~\cite{PhysRevLett.113.208702}.
Since we have by definition
$\hat{p}_c \geq \bar{p}_c$,
it is not a big surprise to find that
$\hat{p}_c$ generates always better
predictions than $\bar{p}_c$~\cite{PhysRevLett.113.208702}.
It is however worth to remark that, whereas in synthetic networks
the difference between the inverse
of the largest eigenvalue of the adjacency
matrix $\bar{p}_c$ and $\hat{p}_c$ can be very large,
in real networks, the two indicators generate pretty
similar estimations of the percolation
threshold, suggesting a relatively little
advantage in using $\hat{p}_c$
in place of $\bar{p}_c$ (see Supplemental Material).
It is interesting to
observe that, in about the $40\%$ of the real networks
analyzed, the naive estimator $\tilde{p}_c$, based only
on the fraction between first and second moments on the
degree distribution, outperforms $\hat{p}_c$
in spite of using much less topological information.
On the other hand, the use of $\tilde{p}_c$ as
prediction tool has the disadvantage
of alternatively over- or under- estimate
the value of the percolation threshold.
The relative error of $\hat{p}_c$
is an increasing function of the
percolation threshold $p_c$.
$\tilde{p}_c$ approximates better $p_c$
as the percolation threshold increases.
Also, the performances of the indicators
do not seem to be strongly related
to the edge density of graph (see Supplemental
Material). Overall, the predictive power
of $\tilde{p}_c$ is lower than the one
of $\hat{p}_c$, as $\tilde{p}_c$ often
predicts too low values of the percolation
threshold.

\begin{figure}[!htb]
\includegraphics[width=0.48\textwidth]{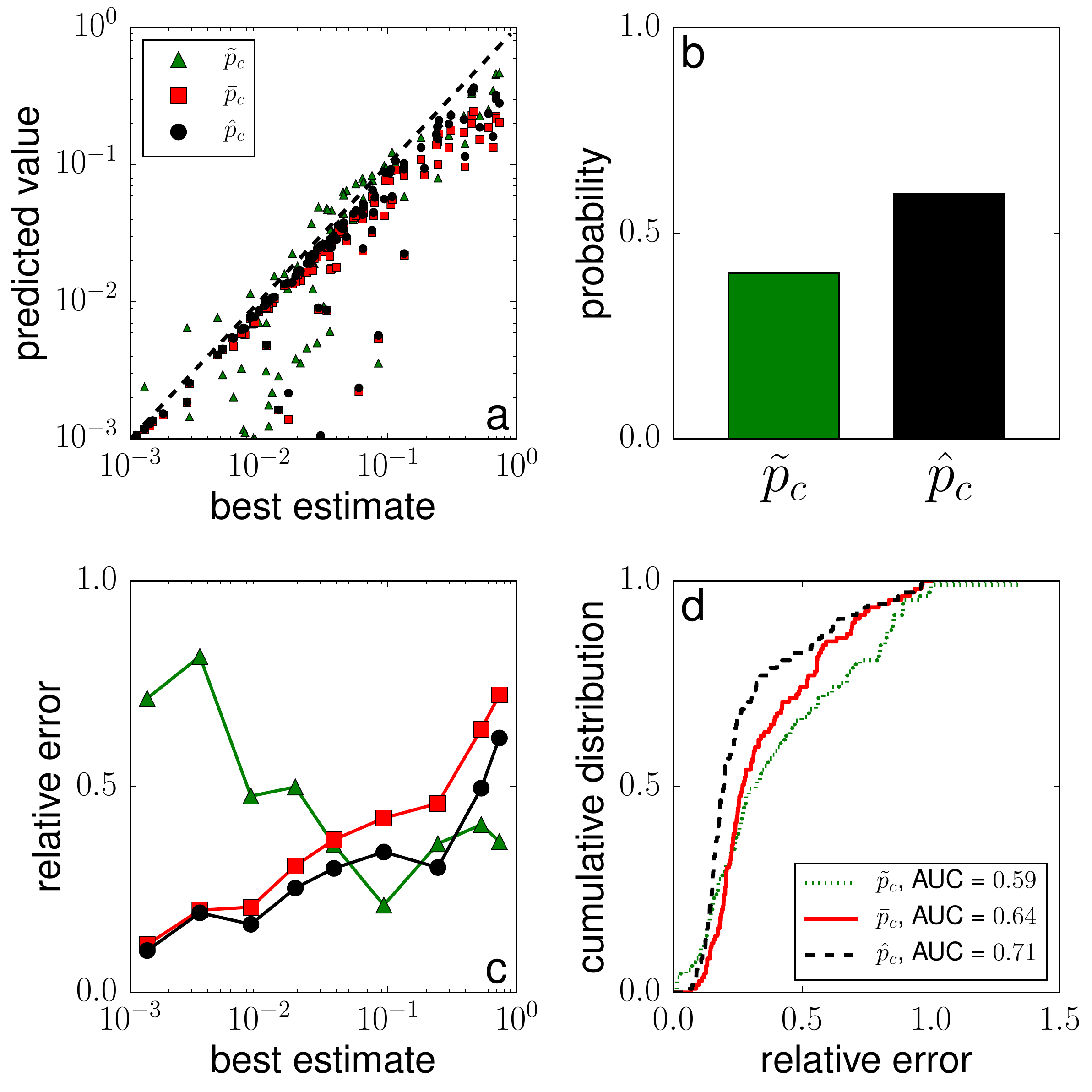}
\caption{
  (Color online) Comparison between best estimates and predicted
  values of the percolation threshold of $109$ real
  networks. In each network, the
  best estimate $p_c$ of the percolation threshold
  [Eq.~(\ref{eq:critical})]
  is computed with $Q=10,000$ independent realizations
  of the Montecarlo method
  by Newman and Ziff~\cite{newman2000efficient}. a) For each network, we
  plot [Eq.~(\ref{eq:moments}), green triangles],
  $\bar{p}_c$ [Eq.~(\ref{eq:adj}, red squares],
  and $\hat{p}_c$ [Eq.~(\ref{eq:non}), black circles]
  as functions of the best estimate $p_c$.
  The black dashed line represents perfect agreement
  between predicted values and best estimates.
  b) Probability, computed
  over the entire sample of real networks,
  that either $\tilde{p}_c$ (green)
  or $\hat{p}_c$ assume the closest value
  to $p_c$.
  c) Relative error of the three predictors,
  with respect to the value of the best estimate,
  as function of $p_c$. We use the same symbols
  and colors as those of panel a.
  d) Cumulative distribution of the relative
  error between predictions and best estimates.
  Green dotted line refers to $\tilde{p}_c$,
  red full line to $\bar{p}_c$, and black dashed line to
  $\hat{p}_c$. Global performances of the prediction
  methods are measured in terms of the area
  under the curve (AUC).
}
\label{fig:3}
\end{figure}

\begin{figure*}[!htb]
\includegraphics[width=0.76\textwidth]{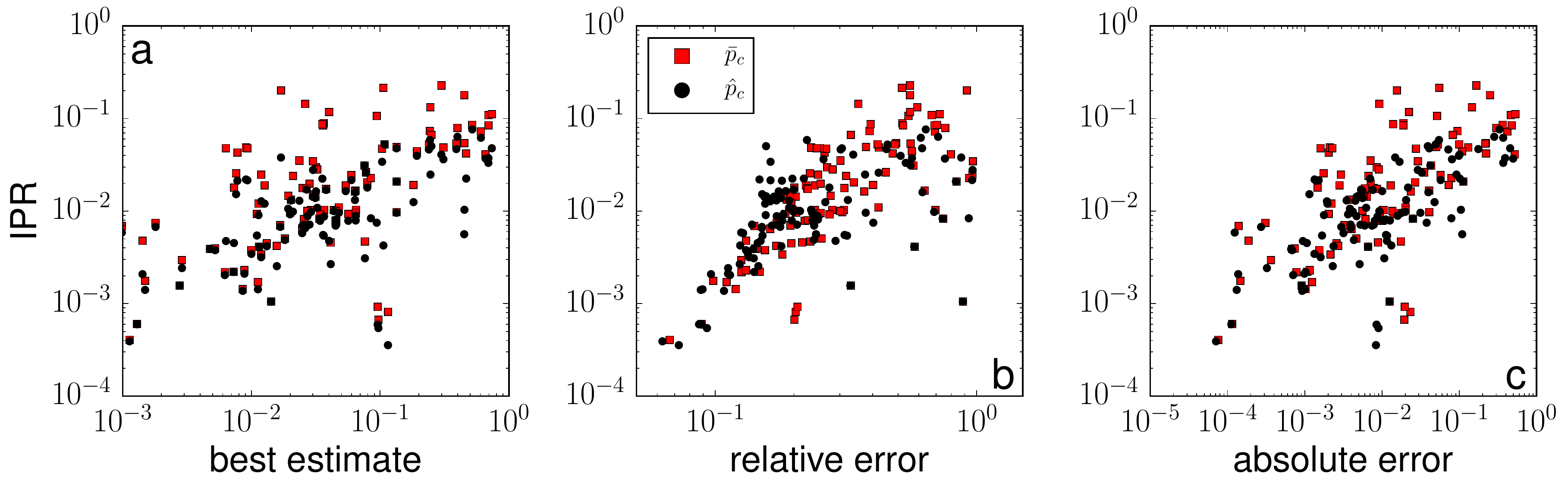}
\caption{(Color online) Inverse participation ratio (IPR) of the principal
  eigenvectors of the adjacency 
  (red squares) and the non-backtracking
  matrices [Eq.~\ref{eq:non_matrix}, black circles]
  as functions of the value of the best estimate
  of the percolation threshold $p_c$ (panel a),
  the relative error (panel b), and the absolute error (panel c).
  Each point refers
  to numerical results obtained on one of the $109$ real
  networks considered in our analysis.
}
\label{fig:4}
\end{figure*}

On the basis of our results, we can
conclude that the inverse
of the largest eigenvalue of the
non-backtracking matrix $\hat{p}_c$ represents the
best measure currently available on the market, among
those that do not rely on direct numerical simulations
of the percolation process,
to provide estimates of the bond percolation threshold
$p_c$ in an arbitrary network. Apparently,
its predictive power is not much influenced
by the density of edges, as long as the network is
sufficiently sparse. Performances are instead
strongly dependent on the value of the
true percolation threshold.
$\hat{p}_c$ generates particularly good
prediction values for bond percolation
thresholds in networks, if
threshold values are sufficiently small.
On the other hand, it fails
by non negligible amounts, if the true
percolation threshold is large. This fact
substantially happens for infrastructural networks, such as
road networks~\cite{leskovec2009community} and
power grids~\cite{watts1998collective},
off-line social networks~\cite{zachary1977information,
  lusseau2003bottlenose, milo2004superfamilies}, and biological
networks~\cite{milo2004superfamilies}
(see Supplemental Material).
Also, this fact often happens in networks with
double percolation transitions~\cite{colomer},
where $\hat{p}_c$ tends to predict the lower
threshold that does not necessarily coincide
with the largest jump of the percolation
strength.

The inadequacy of the indicator
$\hat{p}_c$ to provide good predictions
for $p_c$ in the regime of large
percolation thresholds may be related to
the localization of the eigenvector
associated with the
principal eigenvalue of the
matrix defined in Eq.~(\ref{eq:non_matrix}).
Since the second $N$ components of any eigenvector
of the matrix $M$ are proportional to the first
$N$, we quantify the localization of such eigenvector
by normalizing the first $N$ components (i.e.,
the sum of their squares equals one), and then
computing its inverse participation ratio,
defined as the sum of the fourth power of
the components.
Our measures performed on real networks
indicate the presence of a positive correlation between
the value of the inverse participation ratio
and the value of the best estimate of
the percolation threshold
$p_c$ (Fig.\ref{fig:4}a). They
also show a positive correlation between
the inverse participation ratio
and the relative and absolute errors of $\hat{p}_c$
with respect to $p_c$ (Figs.\ref{fig:4}b and c).
Thus the reason of the bad performance
of $\hat{p}_c$ seems analogous to the
one valid for spectral estimators 
of epidemic thresholds in real networks~\cite{PhysRevLett.109.128702}.
Mathematically speaking
when the principal
eigenvector is localized, the percolation transition
placed at $\hat{p}_c$ involves in reality
only a finite fraction of the nodes in the network, and thus
does not correspond to the true percolation transition
located instead at $p_c$. The same type of considerations
stressed for the principal eigenvector
of the non-backtracking matrix
can be also extended to the principal eigenvector of the
adjacency matrix to motivate why $\bar{p}_c$ suffers from the
same limitations as those observed for $\hat{p}_c$
(Fig.~\ref{fig:4}).
 
As a final consideration, we would like to stress that
the existence of a positive relation
between the error committed by $\hat{p}_c$ to
predict $p_c$ and the value of $p_c$ itself
is an important limitation
for the use of $\hat{p}_c$ as a proxy
for network robustness.
In this context, percolation is better 
thought by removing
than adding edges, so that the robustness
of a network is given by $1-p_c$. The prediction
$1 - \hat{p}_c$ always overestimates
the true fraction of edge failures
that a system can tolerate, providing therefore
an information not useful to prevent 
catastrophic events.
Even more importantly, 
in the analysis of network
resilience to random attacks, it is generally
more meaningful to consider
site instead of bond percolation.
Site percolation thresholds are
larger than those valid for the
edges (see Supplemental Material), and
$1 - \hat{p}_c$ seems not able
to provide sufficiently accurate predictions
for the maximal number of
vertices that can be removed from a network
before reaching system failure.
For all these reasons, we believe that 
additional theoretical efforts must be devoted to
the search of good predictors able to determine
tight upper bounds of percolation 
thresholds in networks.

\begin{acknowledgements}
  The author thanks M.~Bogu\~na and S.~Colomer-de-Sim\'on
  for comments and suggestions. The author is indebted
  to A.~Flammini and C.~Castellano for fundamental
  discussions on the subject of this article.
\end{acknowledgements}


\end{document}